\documentclass[10pt,letterpaper,twocolumn]{article} 

\usepackage{ol2}  

\usepackage{amsmath,epsfig}

\begin{document}

\twocolumn[ 

\title{Broadband Optical Serrodyne Frequency Shifting}


\author{D. M. S. Johnson$^{*}$, J. M. Hogan, S.-w. Chiow and M. A. Kasevich}

\address{
Department of Physics, Stanford University, Stanford, California 94305, USA\\
$^*$Corresponding author: david.m.johnson@stanford.edu
}

\begin{abstract}
We demonstrate serrodyne frequency shifting of light from 200 MHz to 1.2 GHz with an efficiency of better than 60 percent.  The frequency shift is imparted by an electro-optic phase modulator driven by a high-frequency, high-fidelity sawtooth waveform that is passively generated by a commercially available Non-Linear Transmission Line (NLTL).  We also implement a push-pull configuration using two serrodyne-driven phase modulators allowing for continuous tuning between -1.6 GHz and +1.6 GHz.  Compared to competing technologies, this technique is simple and robust, and offers the largest available tuning range in this frequency band.
\end{abstract}

\ocis{000.2170, 230.2090, 230.4110.}
 ] 


\noindent Optical frequency shifting has broad commercial and scientific applications.  For example, optical telecommunication FSK protocols can be implemented using single-sideband (SSB) frequency shifters\cite{Kawanishi2007}.  In spectroscopy and laser cooling experiments, agile frequency shifts from the MHz through the GHz range are often required\cite{Hall1981,Chu1985}.

Acousto-optic modulators (AOMs) are commonly used for MHz-level shifts, but a given AOM has a limited tuning range which is only a small fraction of its fixed center frequency.  AOMs that can deliver GHz-level shifts typically offer this increased bandwidth by sacrificing diffraction efficiency.  Broadband electro-optic phase modulators avoid many of the bandwidth and tuning range limitations of AOMs.  Commercially available fiber phase modulators can have modulation bandwidths above $10~\text{GHz}$.  However, typical sinusoidal phase modulation is an inherently inefficient method of frequency shifting. At best, the fraction of the power in the first-order sideband is limited to $\eta = (J_1(\beta_\text{max}))^2\approx 0.34$, where $J_n(\beta)$ are the Bessel functions of the first kind.  Also, the presence of large undesirable frequency spurs at other harmonics can cause problems in some applications.

In this Letter, we use a serrodyne phase modulation signal\cite{Wright1982} to demonstrate broadband electro-optic frequency shifting with high efficiency $(60 - 80\%)$ into the desired sideband and correspondingly small undesirable spurs.

We briefly review the theory of serrodyne phase modulation\cite{Johnson1988}.  The electric field for the light exiting the phase modulator is $E(t) = E_0 \cos{\left(\omega t + \phi(t)\right)}$ where $\omega$ is the optical frequency and $\phi(t)$ is the phase imprinted by the modulator.  A direct frequency shift can be imparted by applying a linear phase ramp,
\begin{equation}\phi(t) = m\,\delta\, t \!\!\!\!\mod{2\pi m}\label{Eq:SerrodynePhaseShift}\end{equation}
where $\dot{\phi}=m\delta$ is the desired frequency shift and $m$ is an integer.  This phase ramp need only extend from $0$ to $2\pi m$ before resetting since cosine is $2\pi$-periodic.  The resulting serrodyne waveform is a sawtooth with angular frequency $\delta$ and amplitude $2\pi m$.  The first order serrodyne shift occurs for $m=1$ and results in a frequency shift $\delta$, but higher order shifts $(m>1)$ are also possible.  Although a serrodyne signal can theoretically mimic a linear phase ramp within a finite tuning range, it requires a high bandwidth to faithfully reproduce the discontinuities at the end of each period.

The usefulness of the serrodyne technique is limited by the quality of the sawtooth waveform that can be produced within an experimentally accessible bandwidth, and this gets increasingly difficult for higher frequency shifts.  Early work was done with MHz-level frequency shifts\cite{Wright1982,Thylen1985}, and more recently improved undesirable sideband suppression has been demonstrated at these frequencies\cite{Johnson1988,Laskoskie1989,Ozharar2005,Ozharar2007}. Higher frequency serrodyne shifts have been achieved with arbitrary waveform synthesis using mode-locked lasers\cite{Shen2004,Poberezhskiy2005}, but these techniques are complicated and have limited frequency tunability.

In this work we generate high frequency ($200~\text{MHz}$ - $1.6~\text{GHz}$) sawtooth waveforms with good fidelity by using a Non-Linear Transmission Line (NLTL).  In an NLTL, an electronic signal experiences an amplitude-dependent propagation speed\cite{Rodwell1994}.  This effect results in a steepening of the input waveform as the higher amplitude components catch up with the lower amplitude components.  A sinusoidally driven NLTL will therefore output an approximate sawtooth waveform at the drive frequency.  The NLTLs that we use are commercially available, passive components and generate harmonic content out to greater than $20~\text{GHz}$.

\begin{figure}[htb]
\centerline{
\includegraphics[width=8.3cm]{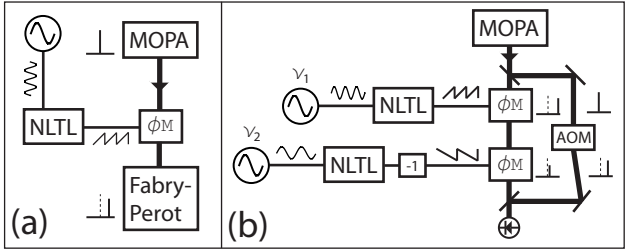}}
 \caption{\label{Fig:Setup} Experimental setup. (a)  A single phase modulator ($\phi M$) measured with a Fabry-Perot cavity.  (b) Two phase modulators in a push-pull configuration measured via a heterodyne beat with the reference beam shifted $194~\text{MHz}$ by an AOM.}
\end{figure}

We directly drive a $\text{LiNbO}_3$ fiber phase modulator (Photline Technologies NIR-MPX850-LN08, $>8 ~\text{GHz}$ bandwidth) with the serrodyne signal generated by an NLTL and analyze the resulting spectrum with a Coherent model 240 Fabry-Perot spectrometer (Fig. \ref{Fig:Setup}a).  We drive the NLTLs by amplifying the output of a signal generator (HP83712A) with a broadband RF amplifier (Mini-Circuits ZHL-42W).  The NLTLs produce high output power serrodyne signals, and since the fiber phase modulators have a low $V_{\pi}\simeq 8~\text{V}$, we can directly drive the modulator without post-amplifying, and thus further bandwidth limiting, the NLTL output.

We generate $780 ~\text{nm}$ narrow-linewidth ($\sim 1 ~\text{MHz}$) laser light using a MOPA laser setup (New Focus Vortex injecting an Eagleyard Tapered Amplifier). A small fraction of the light ($30 ~\text{mW}$) is coupled into the fiber phase modulator, providing up to $14 ~\text{mW}$ of output light.  We couple a fraction of this power into the Fabry-Perot cavity (FSR $7.5 ~\text{GHz}$, resolution $\sim 25 ~\text{MHz}$).

\begin{figure}[htb]
\centerline{
\includegraphics[width=8.3cm]{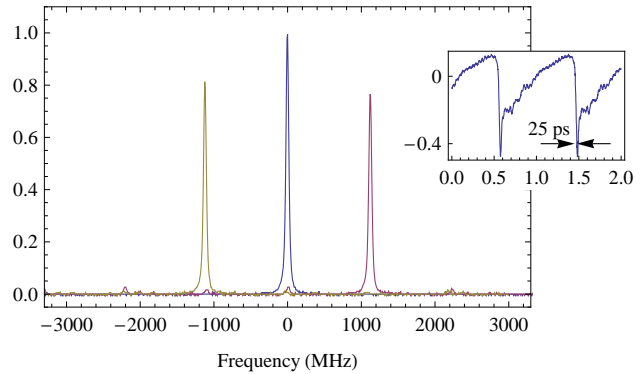}}
 \caption{\label{Fig:FabryPerot} Optical spectrum with and without serrodyne modulation normalized to the unmodulated carrier.  The center (blue) curve shows the optical spectrum with no RF modulation.  The right (red) and left (yellow) curves are the spectrum with a serrodyne modulation of +1.1 GHz and -1.1 GHz, respectively.  The inset shows the applied 1.1 GHz serrodyne waveform in volts versus time in nanoseconds as measured by a 63 GHz oscilloscope.}
\end{figure}

The serrodyne output of the NLTL varies with input frequency and amplitude.  We characterized three NLTLs manufactured by Picosecond Pulse Labs with different frequency ranges and rated input powers: 7112-110 ($300-700 ~\text{MHz}$ @ 29 dBm), 7113-110 ($600-1600 ~\text{MHz}$ @ 29 dBm) and 7102-110 ($300-700~\text{MHz}$ @ 24 dBm, requiring a ZHL-42W post-amplifier).  Each NLTL was swept through its frequency range, and the input amplitude was varied at each point to determine the optimum serrodyne signal.  We captured a data spectrum and a reference spectrum (no RF drive applied to the NLTL) at each drive frequency with a digital oscilloscope.  Example traces are shown in Fig. \ref{Fig:FabryPerot}.  Note that since the fiber phase modulator is electrically floating, we can invert the signal and ground connections to apply a minus sign to the serrodyne signal (a high frequency balun could also be used).

\begin{figure}[htb]
\centerline{
\includegraphics[width=8.3cm]{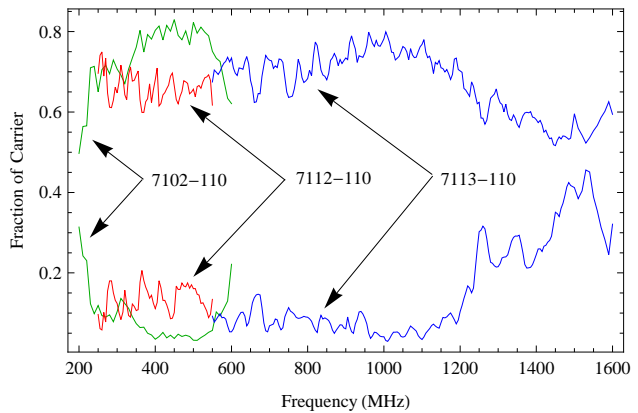}}
 \caption{\label{Fig:ModulationEfficiency} Serrodyne modulation efficiency.  The upper and lower curves show $\eta$ and $\text{SF}$, respectively.}
\end{figure}

The efficiency $\eta$ of the serrodyne is the fraction of the unmodulated carrier power shifted into the desired sideband.  Our serrodyne frequency shift has $\eta > 0.6$ from $200 ~\text{MHz}$ to $1.2 ~\text{GHz}$ as shown in Fig. \ref{Fig:ModulationEfficiency}.  Of particular interest are the regions from $400-500 ~\text{MHz}$ and from $1-1.1 ~\text{GHz}$ which maintain $\eta \sim 0.8$. This is comparable to a well-aligned single pass AOM.  Similarly, we also characterize the cleanliness of the resulting spectrum by its spurious sideband fraction $\text{SF}$, defined as the ratio of the largest spurious frequency component to the desired signal. We measure $\text{SF} < 0.2$ from $200 ~\text{MHz}$ to $1.2 ~\text{GHz}$ as shown in Fig. \ref{Fig:ModulationEfficiency}.  In the region from $700 ~\text{MHz}$ to $1.1 ~\text{GHz}$ we find $\text{SF} < 0.1$.  Figure \ref{Fig:FabryPerot} (left trace) shows a shift with $\eta=0.82$ and $\text{SF}=-16~\text{dB}$.

As a consistency check of the observed serrodyne spectra, we made a time domain measurement of the applied sawtooth waveform $V(t)$ using a 63 GHz Agilent Infiniium DCA 86100B oscilloscope (Fig. \ref{Fig:FabryPerot} inset).   We then calculated the implied serrodyne spectrum using $\phi(t)=\alpha V(t)$ and fit the result to the observed spectrum shown in Fig. \ref{Fig:FabryPerot} with a single free scale parameter $\alpha$.  The predicted spectrum matches the observations very well, with an RMS peak height difference of less than $2\%$ relative to the carrier.

In addition to the first order serrodyne results, we demonstrated second order serrodyne shifts ($m=2$ in Eq. \ref{Eq:SerrodynePhaseShift}) by increasing the amplitude of the sawtooth waveform using the post-amplified 7102-110.  This allows for a larger serrodyne frequency shift of 2$\delta$ without exceeding the frequency range of the NLTL.  We observed second order shifts with $\eta_{2} > 0.5$ and $\text{SF}_2 < 0.25$ for a frequency shift range of $800 ~\text{MHz}$ to $1~\text{GHz}$.  We suspect that these results were limited by the $4.2 ~\text{GHz}$ bandwidth of the post-amplifier.

\begin{figure}[htb]
\centerline{
\includegraphics[width=8.3cm]{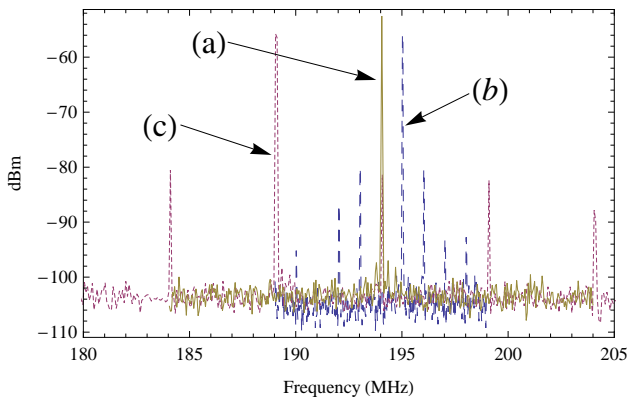}}
 \caption{\label{Fig:pushPullSerrodyne} Heterodyne spectra of push-pull frequency shift. (a) No applied serrodyne modulation signal. (b) $\nu_1=600~\text{MHz}$, $\nu_2=599~\text{MHz}$ and $\Delta\nu_1=+1~\text{MHz}$.  (c) $\nu_1=600~\text{MHz}$, $\nu_2=605~\text{MHz}$ and $\Delta\nu_1=-5~\text{MHz}$.}
\end{figure}

To obtain smaller frequency shifts, we arranged two identical fiber phase modulators in a push-pull configuration (Fig. \ref{Fig:Setup}b).  The two modulators are driven by NLTLs supplied by independently tunable amplified function generators running at frequencies $\nu_1$ and $\nu_2$, respectively.  With this flexible setup we can continuously scan the serrodyne output at $\Delta \nu = \nu_1-\nu_2$ from positive to negative frequencies out to the maximum frequencies of the NLTLs.  Notice that this scheme allows for very small frequency shifts (all the way to zero) that would not be possible with an AOM.  We used a heterodyne measurement to characterize the push-pull spectrum for $\Delta\nu$ smaller than the linewidth of the Fabry-Perot (Fig. \ref{Fig:pushPullSerrodyne}).  An AOM in the reference arm of the interferometer shifted the carrier by a fixed frequency of $194~\text{MHz}$ so that we could distinguish positive frequency beat notes from negative frequency beat notes with respect to the unshifted carrier.  We find that higher-order harmonics of $\Delta\nu$ are suppressed by at least $25~\text{dB}$ compared to the desired shifted signal.

We investigated the phase noise of the serrodyne-shifted light by comparing the unmodulated heterodyne beat note signal to that of the push-pull signal out to $1~\text{MHz}$ with a resolution bandwidth of $1~\text{kHz}$.  The phase noise of the unmodulated signal was $-95~\text{dBc}/\text{Hz}$ at 150 kHz offset. During serrodyne modulation we observe no increase in phase noise above this noise floor.

Optical SSB modulation using a dual Mach-Zehnder geometry\cite{Shimotsu2001} is another well-known solution for frequency shifting in the GHz range.  In this configuration the maximum efficiency is limited to $\eta_\text{SSB}=(J_1(\beta_\text{max}))^2\approx 0.34$, at which point the $-3\delta$ spur is only suppressed by $\text{SF}_\text{SSB}=\left(\frac{J_3(\beta_\text{max})}{J_1(\beta_\text{max})}\right)^2\approx -15~\text{dB}$.  Our measured efficiency is better than this limit, and in some frequency ranges the spur suppression is better as well.  Additionally, the NLTL serrodyne scheme needs only a single phase modulator compared to the four that must be integrated into an SSB modulator.  Finally, the multiple path Mach-Zehnder geometry is a potential source of low-frequency drift.

The serrodyne technique can potentially be improved to offer cleaner frequency shifts and a larger tuning range.  The fall time is the principle performance driver of a serrodyne shift\cite{Johnson1988}, and NLTLs are capable of generating sub-picosecond transients\cite{Weide1994}.  This is more than $50$ times faster than our measured fall time.  Additionally, the serrodyne could be improved by adding a passive network to adjust the phase and amplitude of the existing frequency components.

Although the dual modulator push-pull configuration allows for a much larger tuning range, it does so at the expense of an additional modulator.  In principle, the two serrodyne drive signals could be subtracted using a high-frequency balun and then applied to a single modulator.

Serrodyne frequency shifting using NLTLs offers several advantages over traditional optical frequency manipulation.  Unlike an AOM, the serrodyne phase modulator does not spatially shift the beam and is thus immune to temperature induced alignment changes.  An NLTL-driven phase modulator offers a greater than three octave dynamic tuning range, and when two modulators are used in a push-pull configuration we have demonstrated continuous tuning from $-1.6~\text{GHz}$ to $+1.6~\text{GHz}$.  This is more flexible than previous frequency shifting techniques in this band.

Many thanks to the Fejer group for lending us the Infiniium oscilloscope.

\end{document}